\documentclass[
    ,final            
  ]
  {aipproc}
\layoutstyle{6x9}

\newcommand{\micron}{\mbox{$\mu{\rm m}$}}

\def\lesssim{\mathrel{\hbox{\rlap{\hbox{%
 \lower4pt\hbox{$\sim$}}}\hbox{$<$}}}}
\def\gtrsim{\mathrel{\hbox{\rlap{\hbox{%
 \lower4pt\hbox{$\sim$}}}\hbox{$>$}}}}
\def\wig#1{\mathrel{\hbox{\hbox to 0pt{   
  \lower.5ex\hbox{$\sim$}\hss}\raise.4ex\hbox{$#1$}}}}
\newcommand{\arcsec}{\mbox{$^{\prime \prime}$}}

\newcommand{\Lbol}{\mbox{$L_{bol}$}}
\newcommand{\Teff}{\mbox{$T_{\rm eff}$}}
\newcommand{\logg}{\mbox{$\log(g)$}}
\newcommand{\HST}{{\sl HST}}

\newcommand{\Mjup}{\mbox{M$_{\rm Jup}$}}
\newcommand{\twomassbin}{\hbox{2MASS~J1534$-$2952AB}}
\newcommand{\hdbin}{\hbox{HD~130948BC}}
\newcommand{\hdprim}{\hbox{HD~130948A}}
\newcommand{\hdage}{\hbox{0.79$^{+0.22}_{-0.15}$~Gyr}}
\newcommand{\Msun}{\hbox{M$_{\odot}$}}

\begin{document}

\title{Fundamental Properties of Low-Mass Stars and Brown Dwarfs}

\author{Michael C. Liu}{
  address={Institute for Astronomy, University of Hawai`i, 2680 Woodlawn Drive, Honolulu, HI 96822}}

\author{Keivan G. Stassun}{
  address={Vanderbilt University, Physics \& Astronomy Department, Nashville, TN 37235 USA}}

\author{France Allard}{
  address={Centre de Recherche Astrophysique de Lyon, UMR 5574: CNRS, Universit\'{e} de Lyon, \'{E}cole Normale Sup\'{e}rieure de Lyon, 46 all\'{e}e d'Italie, 69364 Lyon Cedex 07}}

\author{Cullen H. Blake}{
  address={Harvard-Smithsonian Center for Astrophysics, 60 Garden Street, Cambridge, MA, 02138}}

\author{M. Bonnefoy}{
  address={Laboratoire d'Astrophysique de Grenoble, BP 53, F-38041 GRENOBLE C\'{e}dex 9, France}}

\author{Ann Marie Cody}{
  address={California Institute of Technology, Department of Astrophysics,
  Pasadena, CA 91125 USA}}

\author{A. C. Day-Jones}{
  address={Centre for Astrophysics Research, University of Hertfordshire, Hatfield, AL10 9AB, UK}}

\author{Trent J. Dupuy}{
  address={Institute for Astronomy, University of Hawai`i, 2680
  Woodlawn Drive, Honolulu, HI 96822}}

\author{Adam Kraus}{
  address={California Institute of Technology, Department of Astrophysics,
  Pasadena, CA 91125 USA}}

\author{Mercedes L\'{o}pez-Morales}{
  address={Carnegie Institution of Washington, Dept.\ of Terrestrial Magnetism, 
  Washington, DC 20015 USA}}

\begin{abstract}
  Precise measurements of the fundamental properties of low-mass stars
  and brown dwarfs are key to understanding the physics underlying
  their formation and evolution.  While there has been great progress
  over the last decade in studying the bulk spectrophotometric
  properties of low-mass objects, direct determination of their
  masses, radii, and temperatures have been very sparse.  Thus,
  theoretical predictions of low-mass evolution and ultracool
  atmospheres remain to be rigorously tested.  The situation is
  alarming given that such models are widely used, from the
  determination of the low-mass end of the initial mass function to
  the characterization of exoplanets.

  An increasing number of mass, radius, and age determinations are
  placing critical constraints on the physics of low-mass objects.  A
  wide variety of approaches are being pursued, including eclipsing
  binary studies, astrometric-spectroscopic orbital solutions,
  interferometry, and characterization of benchmark systems.  In
  parallel, many more systems suitable for concerted study are now
  being found, thanks to new capabilities spanning both the very
  widest (all-sky surveys) and very narrowest (diffraction-limited
  adaptive optics) areas of the sky.  This Cool Stars 15 splinter
  session highlighted the current successes and limitations of this
  rapidly growing area of precision astrophysics.

%
%

\end{abstract}



\maketitle

\section{Visual Binaries}

About 100~ultracool (spectral type M6 or later) visual binaries are
known, the product of several major high angular resolution imaging
surveys conducted by \HST\ and ground-based adaptive optics (AO)
imaging (e.g., \cite{2001AJ....121..489R, 2003AJ....126.1526B,
  2003ApJ...586..512B, 2003ApJ...587..407C, 2005astro.ph..8082L,
  2006astro.ph..5037L, 2008AJ....135..580R}).  Most of these belong to
the field population, near enough to Earth to be well-resolved and
many have orbital periods amenable to dynamical mass determinations.
A fundamental characteristic of field objects is that they span a
range of (largely unknown) ages.  This is a particularly important
issue for brown dwarfs, which continually cool in time and thus follow
a mass-luminosity-age relation.  Despite this uncertainty, these
objects can strongly test theoretical models when analyzed
appropriately.

\begin{figure}
  \hskip -0.3in
  \includegraphics[height=.7\textheight, angle=90]{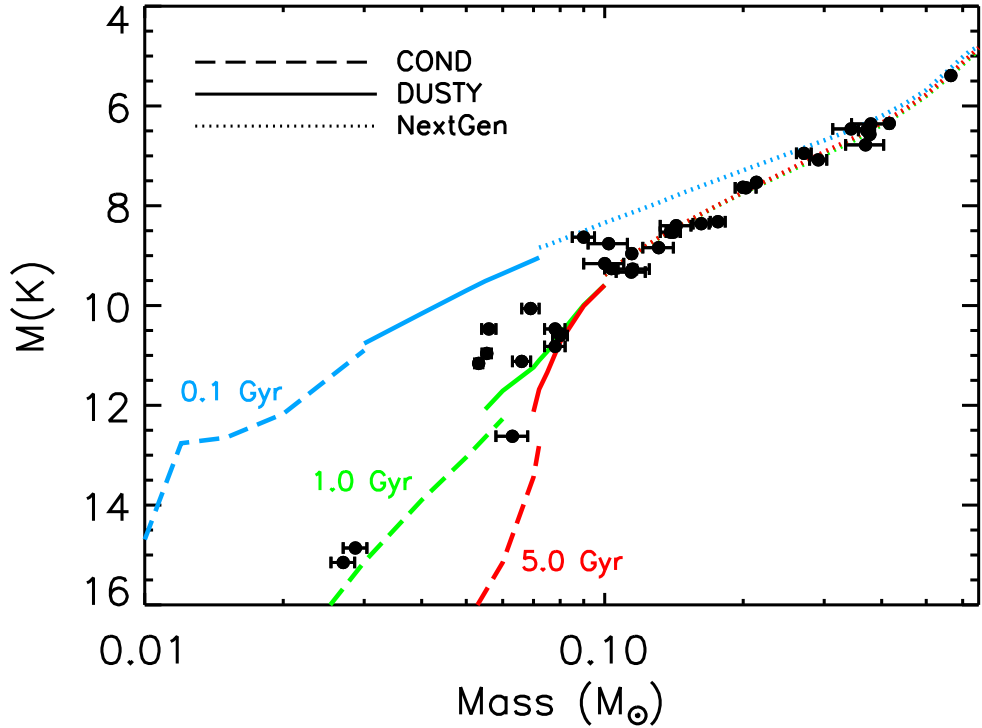}
  \caption{Absolute $K$-band magnitude as a function of dynamical mass
    for field M, L, and T~dwarfs, based on data from
    \cite{2000A&A...364..217D, 2007ApJ...661..496M, gl802b-ireland,
      liu08-2m1534orbit, 2008arXiv0807.2450D, 2004A&A...423..341B,
      2008A&A...484..429S, 2004astro.ph..7334O, 2007ApJ...665..736C}.
    For binaries where only the total mass is measured, the masses of
    the individual components are determined from the observed
    $K$-band flux ratio and evolutionary models to determine the mass
    ratio of the two components.  The errors in the ordinate are
    comparable or smaller than the plotting symbol.  Evolutionary
    models from \cite{2003A&A...402..701B,1998A&A...337..403B} are
    overplotted, with each class of models (NextGen, DUSTY or COND)
    plotted over the range of \Teff\ appropriate for each
    model. \label{fig:mass-magnitude}}
\end{figure}

Accurate masses from visual binaries require high quality astrometry,
radial velocities, and parallaxes (errors of $\sim$1~mas,
$\sim$1~km/s, $\sim$2\%, respectively).  Also, to compare to models,
independent determinations of \Lbol\ to $\lesssim$10\% (a more
challenging measurement than appreciated at face value, e.g.,
\cite{gol04}) and \Teff\ are needed.  Until this year, only three
objects had dynamical masses that placed them unambiguously below the
substellar limit: the M9 tertiary component of the hierarchical triple
Gl~569 \cite{2001ApJ...560..390L, 2004astro.ph..7334O,
  2006ApJ...644.1183S} and both components of the young M6.5+M6.5
eclipsing binary 2MASS~J05352184$-$0546085 in the Orion Nebula
\cite{2006Natur.440..311S}.  Since many of the first binary surveys
were carried out nearly a decade ago, the next few years should see a
rapid increase in the number of dynamical mass determinations, thereby
extending the mass-magnitude relation by about a factor of 10 in mass
and a factor of 100 in luminosity (Figure~\ref{fig:mass-magnitude}).

Liu and collaborators have recently extended such measurements with
the first dynamical mass for a binary T dwarf, the T5+T5.5 system
\twomassbin\ \cite{liu08-2m1534orbit}.
With a total mass of only of $0.056\pm0.003 M_\odot$ ($59\pm3~
M_{Jup}$), this is the coolest and lowest mass binary with a dynamical
mass to date, as well as the first field binary for which both
components are confirmed to be substellar.
%
%
The H-R diagram positions of the two components of \twomassbin\ are
discrepant with theoretical evolutionary tracks.  While this could
stem from large systematic errors in the luminosities ($\sim$50\%
errors) and/or radii ($\sim$20\% errors) predicted by evolutionary
models, the likely cause is that temperatures of mid-T~dwarfs
determined with model atmospheres are too warm by $\approx$100~K.
In fact, these model atmosphere uncertainties are the current limiting
factor in testing theory using the H-R diagram, not the accuracy of the
mass determinations.
Morever, the prediction of different evolutionary models (e.g., Tucson
and Lyon) are essentially indistinguishable on the H-R diagram.

These limitations imposed by atmospheric models can instead be
circumvented, by using accurate mass and luminosity determinations in
concert with {\em evolutionary} models to very precisely infer
physical parameters for substellar binaries and assuming the systems
are coeval.  In the case of \twomassbin, the formal uncertainties on
the age ($\pm$0.1~Gyr), temperature ($\pm$17~K) and surface gravity
($\pm$0.04~dex) allow for strong points of comparison with other data.
For instance, this approach gives a relatively youthful age for the
system of 0.79$\pm$0.09~Gyr, consistent with its low tangential
velocity relative to other field T~dwarfs.
More generally, low-mass field binaries with dynamical mass
determinations (``mass benchmarks'') can serve as precise reference
points for testing \Teff\ and \logg\ measurements from ultracool
atmosphere models, as good as or even better than single brown dwarfs
with age estimates (``age benchmarks'').
In fact, given the plausible observational uncertainties, mass
benchmarks are likely to provide stronger constraints (by a factor of
$\approx$5) on \logg\ and \Teff\ than age benchmarks, since dynamical
masses can be determined far more accurately than ages for field
stars.


\begin{figure}
  \hskip 0.3in
  \includegraphics[height=.5\textheight, angle=90]{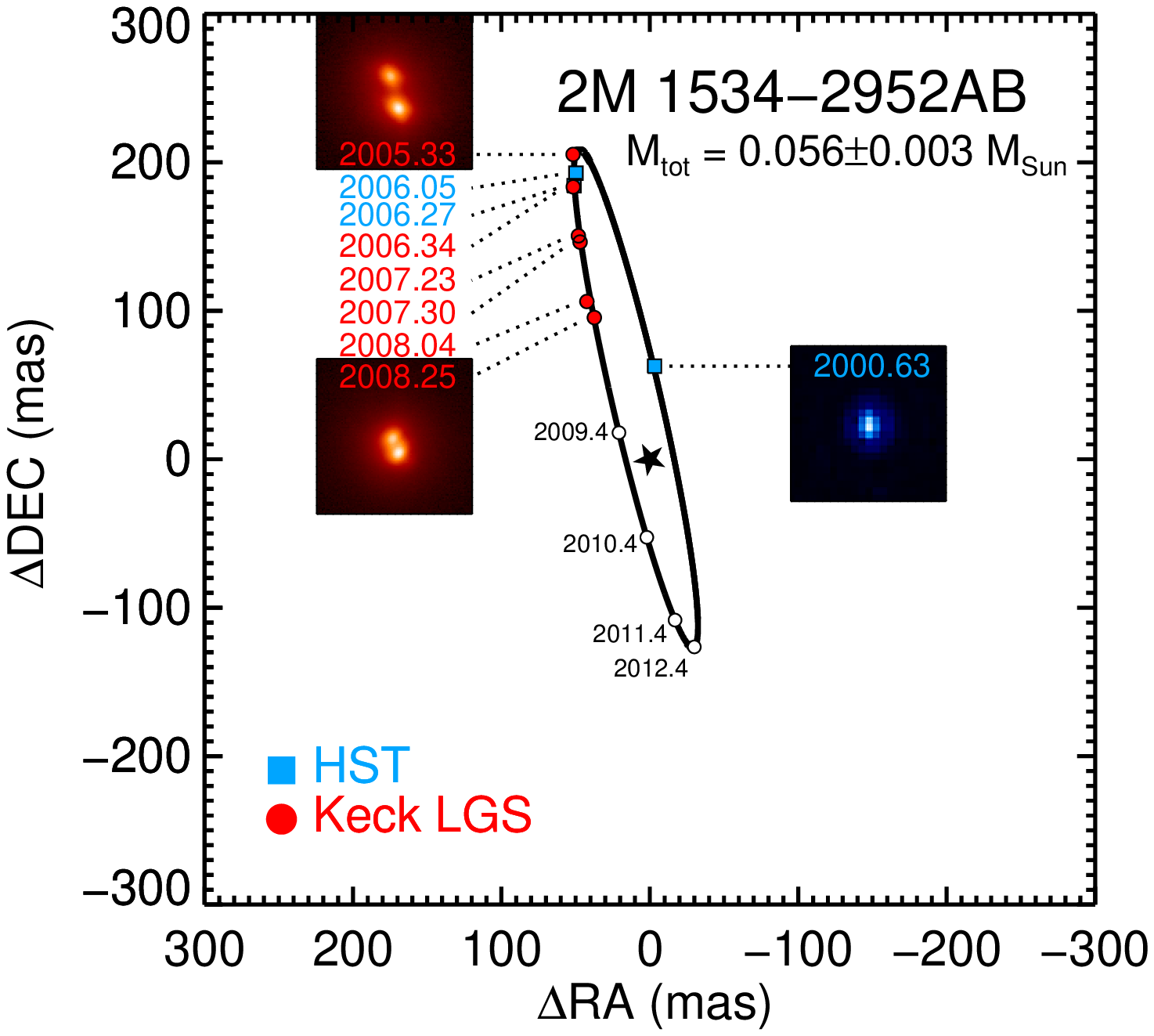}
  \hskip -1.3in
  \includegraphics[height=.5\textheight, angle=90]{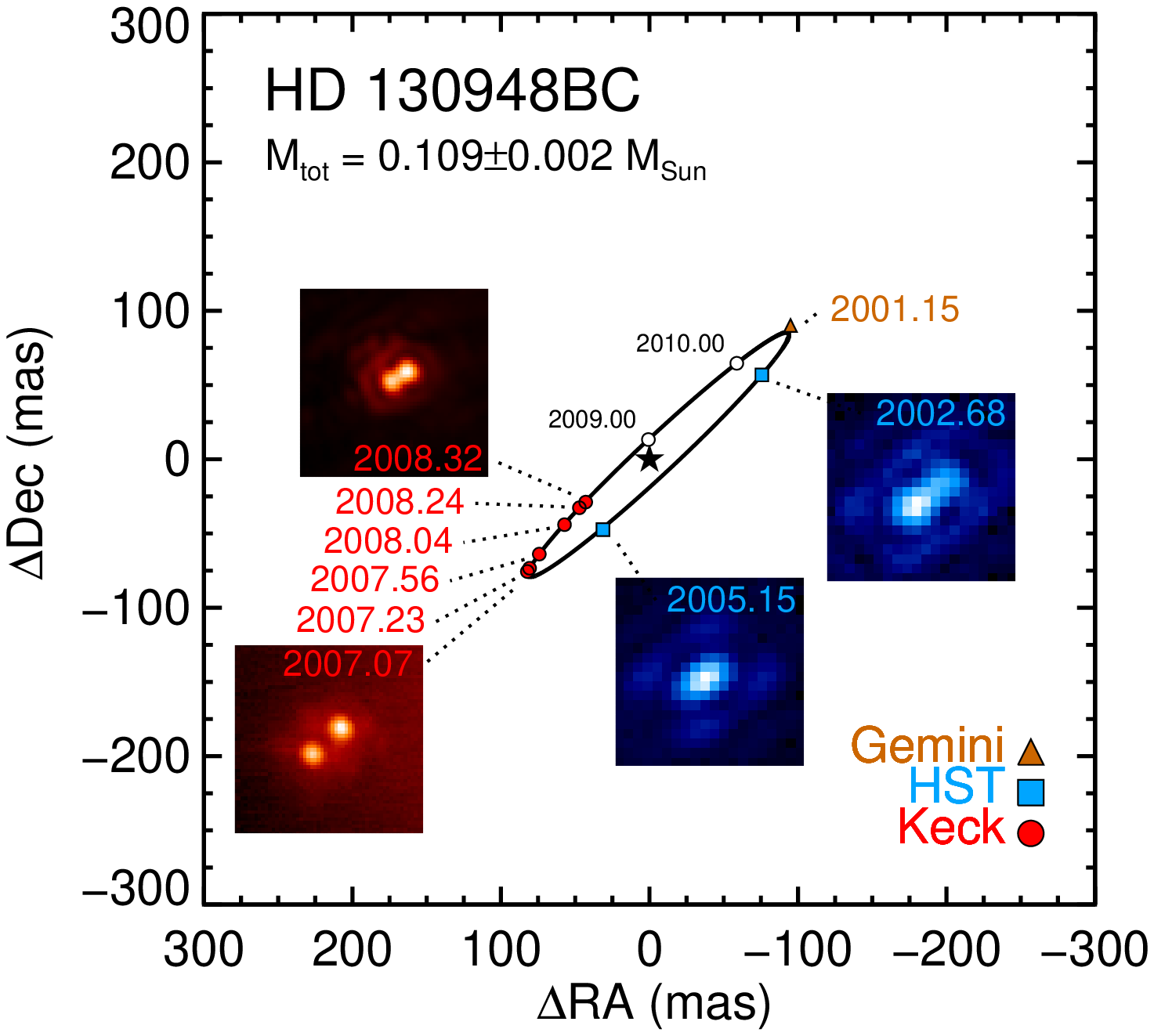}
  \caption{Recent orbit and dynamical mass determinations for field
    brown dwarf binaries.  The insets show imaging data at selected
    epochs and are 1\arcsec\ on a side.  {\em Left:} Orbit for the
    binary T5.0+T5.5 dwarf 2MASS~J1534-2952AB, based on high angular
    resolution monitoring using \HST\ and Keck laser guide star
    adaptive optics \cite{liu08-2m1534orbit}.  This is the coolest and
    lowest mass visual binary to date with a direct mass
    determination. {\em Right:} Orbit for the binary L4+L4 dwarf
    HD~130948BC, based on \HST\ and natural guide star imaging from
    Keck and Gemini \cite{2008arXiv0807.2450D}.  This system also has
    an independent age determination of \hdage, based on the rotation
    and activity properties of its G2V~primary star, making it a thus
    far unique benchmark system for testing theoretical models.}
\end{figure}


Finally, the most stringent tests of theory can come from binaries
with dynamical masses {\em and} independently determined ages, by
being members of star clusters/groups and/or as companions to stars of
known age.  These systems represent the ``gold standard'' for testing
models, but they are very rare.  The splinter session presented new
results for two objects:

\medskip
\noindent $\bullet$ {\em The Young Low-Mass Binary TWA 22AB:}
%
In 2004, Bonnefoy and collaborators resolved the TW~Hydrae Association
member TWA22 as a tight ($\sim 1.8$~AU) binary with the VLT/NACO
instrument.  Follow-up observations have monitored 80\% of the binary
orbit. Armed with the trigonometric parallax of the system, they find
a total dynamical mass (M=220 $\pm$ 21 $M_{Jup}$). Additional
observations with the VLT/SINFONI AO-assisted integral field
spectrometer have obtained medium resolution spectra ($R=1500-2000$)
of the primary and of the companion over the spectral range
$1.0-2.5~\micron$.  Spectral indices, equivalent widths and
least-squares fitting were employed to compare the spectra to
empirical spectral libraries of field and young dwarfs, yielding a
M6V~$\pm$~1 spectral type for both components. Spectral templates were
also used to estimate the temperatures and surface gravities of
TWA22~A and B.  Overall, the measured mass does not agree with 8-Myr
evolutionary tracks. This could mean that TWA22~AB is either older
than expected (and perhaps not a member of TWA) or that models
under-predict masses of young objects.

\medskip
\noindent $\bullet$ {\em The Field Substellar Benchmark Binary HD~130948BC:}
Dupuy and collaborators have been using the Keck AO system to monitor
a large sample of ultracool field binaries to enable a much better
assessment of substellar theoretical models by obtaining many more
dynamical masses.  They presented Keck AO imaging of the L4+L4 binary
\hdbin\ along with archival \HST\ and Gemini-North observations, which
together span $\approx$70\% of the binary's orbital period
\cite{2008arXiv0807.2450D}.  From the relative orbit, they determine a
total dynamical mass of 0.109$\pm$0.002~\Msun\ (114$\pm$2~\Mjup). The
flux ratio of \hdbin\ is near unity, so both components are
unambiguously substellar for any plausible mass ratio. In addition, an
independent constraint on the age of the system is available from the
primary \hdprim\ (G2V, [M/H]~=~0.0). The ensemble of available
indicators suggests an age comparable to the Hyades, with the most
precise age being \hdage\ based on gyrochronology.

As a result, \hdbin\ is now a unique benchmark among field L and
T~dwarfs: it is the only system with a well-determined mass,
luminosity, and age. Thus, the luminosity evolution of brown dwarfs
predicted by theoretical models is fully constrained by observations
for the first time, and the result is that the models disagree with
the data: (1) Both components of \hdbin\ appear to be overluminous by
a factor of $\approx$2--3$\times$ compared to evolutionary models. The
age of the system would have to be notably younger than the gyro age
to ameliorate the luminosity disagreement. (2) Effective temperatures
derived from evolutionary models for HD~130948B and C are inconsistent
with temperatures determined from spectral synthesis for objects of
similar spectral type.  Overall, regardless of the adopted system age,
evolutionary and atmospheric models give inconsistent results, which
indicates systematic errors in at least one class of models, possibly
both.  The masses of \hdbin\ happen to be very near the theoretical
mass limit for lithium burning, such that the Lyon and Tucson models
give drastically different predictions for the lithium depletion that
has occured in each component. Thus, measuring the differential
lithium depletion between B and C will provide a uniquely
discriminating test of theoretical models.

The potential underestimate of luminosities by evolutionary models
would have wide-ranging implications since they are widely used for
characterizing low-mass stars, brown dwarfs, and planets.  Therefore,
a more refined age estimate for \hdprim\ and measurements for more
such mass+age benchmarks are critically needed to determine the
magnitude of the luminosity discrepancy more precisely.

\section{Eclipsing Binaries}

Eclipsing binary (EB) stars generally offer the most accurate means
for directly measuring stellar masses and radii. Of course, EBs are
rare, and thus the mass-radius relation of K and M dwarfs has
historically been poorly constrained.

The situation has improved recently with the discovery of several new
detached EBs, particularly at very low masses and at young ages, as
described in presentations by Stassun, L\'opez-Morales, and Kraus.
For example, in the last few years the number of young ($<30$ Myr),
low-mass ($M<2 {\rm M}_\odot$) EBs has increased to 16 (see
\cite{2007prpl.conf..411M}).  Highlights from this recent work
include: (1) The first EB of equal-mass (``identical twin'') stars
that exhibit striking differences in temperature and luminosity,
suggesting that non-coevality of $\sim 30$\% is possible in young
binaries \cite{2008Natur.453.1079S}, and (2) the first EB system of
two brown dwarfs \cite{2006Natur.440..311S, 2007ApJ...664.1154S} in
which strong magnetic activity is likely responsible for the
surprising reversal of temperatures in the system (the higher mass
brown dwarf is the cooler of the pair). Indeed, there is now growing
evidence that magnetic activity may be affecting the structure, and
thus the basic mass-luminosity relationship, of young, low-mass stars.

Among field stars, the results from EBs combined with stellar radii of
single K and M dwarfs derived via interferometry (in which case the
stellar masses are derived from models) currently add up to over forty
radius measurements that can test stellar models. A graphical summary
is shown in Fig.\ 2, which plots all the available mass and radius
measurements for stars with $M < 1 M_{\odot}$ versus the best fitting
model from \cite{1998A&A...337..403B}.  The figure clearly illustrates
that the radii of many of the stars are significantly larger than
predicted by the models for stars with $M > 0.35 M_{\odot}$, and there
also is significant scatter in the stellar radii. For stars with $M <
0.35 M_{\odot}$, which coincides with stars becoming fully convective,
models and observations seem to agree (but see also below).

\begin{figure}
  \includegraphics[height=.5\textheight, angle=0]{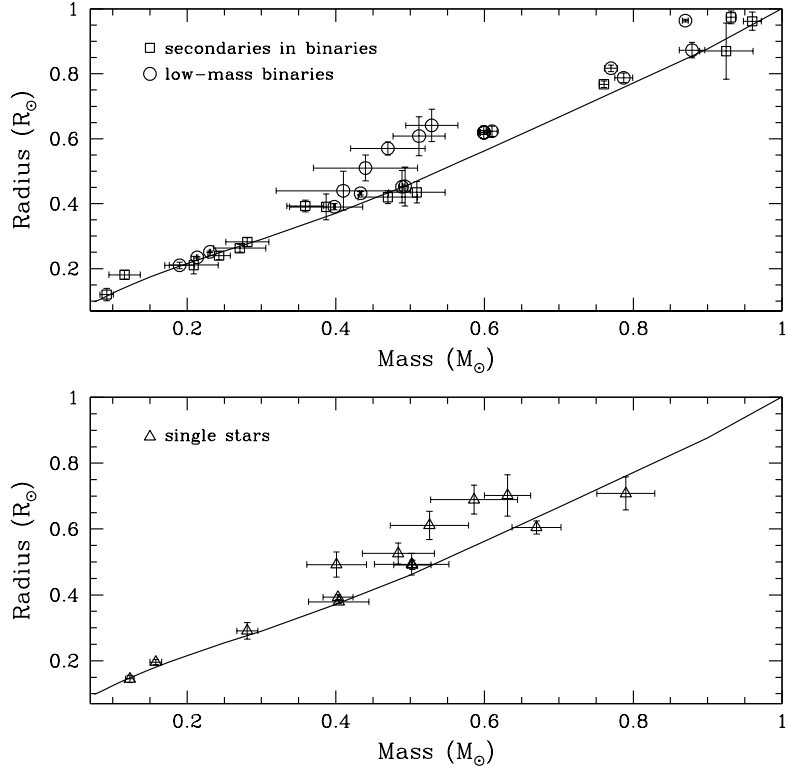}
  \caption{Current observational mass-radius relation for stars below
    $1 {\rm M}_\odot$.  Top, all the data from low-mass secondaries to
    eclipsing binaries with primaries more massive than $1 {\rm
      M}_\odot$ (squares) and the components of eclipsing binaries
    below $1 {\rm M}_\odot$ (circles); bottom, all the measurements
    from single stars. The solid line in each panel represents the
    theoretical isochrone model from \cite{1998A&A...337..403B}, for
    an age of 1 Gyr, Z = 0.02, and mixing length $\alpha = 1.0$
    (standard model). Figure reproduced from
    \cite{2007ApJ...660..732L}.}
\end{figure}

The most plausible explanations for these trends are the magnetic
activity in the atmospheres of these stars, or their metallicity
(equation of state effects cannot currently be tested
observationally).  Magnetic activity is clearly affecting the radii of
the stars, at least in the case of the most active components of
binaries (see Fig.\ 2 of \cite{2007ApJ...660..732L}). In the case of
single stars, this radius--magnetic activity correlation is not as
clear. However, we need to address that this conclusion might suffer
from an observational bias, as all the stars with $M < 0.35 M_{\odot}$
in the sample happen to have low activity levels. Metallicity seems to
also be having some effect in the radii of K and M dwarfs stars, with
more metal-rich objects apparently showing larger radii (see Fig.\ 4
of \cite{2007ApJ...660..732L}). However, the still weakly determined
metallicity scale of low-mass stars, and the scatter in the data,
might prove this last correlation spurious.

More radius measurements of low-metallicity low-mass stars are
necesary.  To that end, work is ongoing to increase the sample of
low-mass EBs further. M dwarfs in particular are ubiquitous in the
solar neighborhood, but their fundamental properties are not as well
understood as those of their more massive brethren.  Only $\sim$12
M-dwarf EBs have been identified to date since low-mass stars are
intrinsically faint and shallow variability surveys have only
encompassed a very limited survey volume. Kraus and collaborators are
conducting a program to identify and characterize new M dwarf EBs from
a deep field variability survey (the 1st MOTESS-GNAT survey;
\cite{2007AJ....134.1488K}).  Thus far, they have identified $\sim$25
new M dwarf EBs with spectral types as late as M4; this sample triples
the number of known systems.  They have obtained radial velocity
curves for 18 of these systems with Palomar, Keck, and Hobby-Eberly,
more than doubling the number of precise mass measurements for M dwarf
EBs, and now they are pursuing an ongoing program to obtain multicolor
eclipse light curves in order to measure the component radii for each
system. When complete, this survey will allow for the first systematic
investigation of the fundamental properties of very low-mass stars.

\section{Atmospheric Models}

The atmospheres of very low-mass stars and brown dwarfs are governed
by the formation of molecules and dust grains in a relatively
high-temperature, high-gravity environment compared to laboratory
experiments.  Over the past decades, it has been shown that these
opacities cover all wavelengths of the emerging spectrum, to the point
of locking the peak of the SED around 1.2~\micron\ as \Teff\ decreases
for metal-rich compositions.
The effect of reducing metallicity would therefore be to recover the
reddening of the SED with decreasing \Teff.  However, double metals
deplete more rapidly and thus density-sensitive features such as
hydride absorption bands (MgH and CaH) and the very important
near-infrared collision-induced H$_2$ absorption are revealed and
shape the SED in enhanced-density atmospheres.
The effects of changing surface gravity are relatively more subtle and
consist of atmospheric density changes at constant composition, which
can be compensated by a \Teff\ change in M~dwarfs.  There is therefore
a degeneracy of solutions in determining the gravity and \Teff, and
therefore age and mass, of very low-mass stars and brown dwarfs.

Atomic line profile analysis is complicated by the uncertain molecular
pseudo-continuum background (oscillator strengths are often inaccurate
or missing).  In addition, for T dwarfs an additional complication is
the lack of this pseudo-continuum background, causing the line wings
of atomic lines to shape the SED more than 2000~\AA\ from the line
center, i.e. beyond the validity of classical assumptions for the line
profile modeling (Lorentz profile with van der Waals collisional
broadening).

Allard and collaborators have developed an online tool (web simulator)
for the determination of parameters based on observed colors by
chi-square fitting and also generating isochrones on any model grid,
filter set, and reddening.  Among the challenges for modeling these
atmospheres and therefore improving the determination of their
parameters will be improvement of (1) the molecular opacities and line
broadening of alkali metals (and molecules), and (2) the modeling of
clouds and non-equilibrium chemistry, which both depend on
understanding of mixing induced by convection into the line formation
layers of these atmospheres.

\section{Other Recent Developments}

\noindent {\bf \em An L Dwarf Radial Velocity Survey:} Precise radial
velocities (RVs) of very low-mass stars and brown dwarfs can provide a
wealth of information about the fundamental physical properties of
these objects. Measurements of projected rotational velocities and
systemic velocities, when coupled with proper motions, provide insight
into the dynamical history of this population. In addition,
multi-epoch RV measurements can be used to search for single- and
double-lined binaries. Binaries provide an excellent opportunity to
directly measure the fundamental physical properties of stars and
brown dwarfs and to compare these measurements to theoretical models.

Blake and collaborators have developed a technique for obtaining
precise RVs of low-mass stars and brown dwarfs in the near infrared
and are conducting a magnitude-limited survey of L~dwarfs with the
NIRSPEC spectrograph on the Keck telescope.  The sample consists of 75
L dwarfs with observations spanning up to four years. With a typical
RV precision of 200 m s$^{-1}$, they are very sensitive to low-mass
binaries with orbital separations smaller than those probed by direct
imaging techniques.  They have discovered one new single-lined L dwarf
binary, 2MASS 0320$-$04, likely comprised of a late-M and a T dwarf
(\cite{2008ApJ...678L.125B}; see also \cite{2008arXiv0803.0295B}).  A
more detailed analysis of the RV data for this system may result in
the detection of the spectral lines of the fainter component.  This
would provide a direct measurement of both the mass and luminosity
ratios of the binary components, allowing us to directly test
theoretical models of brown dwarfs.

\bigskip
\noindent {\bf \em Searching for benchmark brown dwarfs as members of
  binary systems:} For the majority of known brown dwarfs, properties
such as gravity and metallicity remain uncertain, and it is not yet
fully understood how factors these affect their spectra or change over
time.  The complex nature of ultracool and brown dwarf atmospheres
leaves models incomplete or with large uncertainties on their
values. What is vitally needed are benchmark objects, where properties
can be determined independently.

Day-Jones and collaborators are currently undertaking a search to find
such benchmarks as members of binary systems containing an
age-calibratable primary.  White dwarf or subgiant primaries can
provide accurate ages, distances and, for subgiants, metallicity
constraints, which can then be applied (by association) to the brown
dwarf companions.  They have searched for widely separated
($\sim$20,000~AU) ultracool+white dwarf binaries in 2MASS and
SuperCOSMOS in the southern hemisphere and have to date discovered one
ultracool+white dwarf binary system \cite{2008MNRAS.388..838D}, which
is the widest separated M9 + WD binary known to date and has an age
constraint of $>$1.94~Gyrs.  To find similar systems, they have also
mined the latest releases of SDSS (DR6) and UKIDSS Large-Area Survey
(DR3) and carried out a pilot imaging survey in the south; these have
yielded a good number of candidates that are currently being followed
up.


\bigskip
\noindent{\bf \em Searching for Pulsation in Brown Dwarfs and Very Low
  Mass Stars:} Censuses of young ($\sim$1-10 Myr) clusters over the
past decade have revealed substantial numbers of very low mass stars
and brown dwarfs.
In order to place more constraints on the physical properties of young
brown dwarfs, Cody and collaborators have begun a photometric campaign
that aims to uncover a pulsational instability on hour timescales in
those objects that are still burning deuterium
\cite{2005A&A...432L..57P}.  The identification of pulsations would
provide a new probe of brown dwarf interiors through the physics of
seismology.  The campaign is ongoing, with completed or planned
observations of some 80~brown dwarfs and very low mass stars in five
young star clusters, using telescopes at the Palomar and CTIO.
Preliminary results in the IC~348 and $\sigma$~Orionis clusters
include several brown dwarfs displaying variability with periods of a
few hours at the limit of detectability.  High-resolution
spectroscopic followup is underway to determine whether rapid
rotational modulation of magnetic spots can explain the light curves,
or if the variability can indeed be attributed to a new class of
pulsation.


\newcommand\aj{AJ}
\newcommand\araa{ARA\&A}
\newcommand\apj{ApJ}
\newcommand\apjl{ApJ}
\newcommand\apjs{ApJS}
\newcommand\aap{A\&A}
\newcommand\aapr{A\&A~Rev.}
\newcommand\aaps{A\&AS}
\newcommand\baas{BAAS}
\newcommand\mnras{MNRAS}
\newcommand\nat{Nature}
\newcommand\pasp{PASP}


%
%
%
%
%


\end{document}